\documentclass[twocolumn,secnumarabic,amssymb, nobibnotes,superscriptaddress,showpacs, prb, footinbib,aps]{revtex4-2}

\usepackage[utf8]{inputenc}
\usepackage{amsmath}
\usepackage{overpic}
\usepackage{caption}
\usepackage{subcaption}
\usepackage{multirow}
\usepackage{enumerate}
\usepackage{comment} 
\usepackage[normalem]{ulem}
\usepackage[colorlinks=true,linkcolor=black,citecolor=blue,urlcolor=blue]{hyperref}

\usepackage{graphicx}
\usepackage{multirow}
\usepackage{dcolumn}
\usepackage{bm}
\usepackage[utf8]{inputenc}
\usepackage[T1]{fontenc}
\usepackage{etoolbox}
\usepackage{xcolor}
\usepackage{soul}
\usepackage{siunitx}

\newcommand{\efield}[1]{$E$\textsubscript{ext}}
\newcommand{\pr}{$P$\textsubscript{r}}{}
\newcommand{\ec}{$E$\textsubscript{c}}{}
\newcommand{\bto}{BaTiO$_3$}{}
{}

\makeatletter
\def\@email#1#2{%
 \endgroup
 \patchcmd{\titleblock@produce}
  {\frontmatter@RRAPformat}
  {\frontmatter@RRAPformat{\produce@RRAP{*#1\href{mailto:#2}{#2}}}\frontmatter@RRAPformat}
  {}{}
}%
\makeatother
\begin{document}

\title[]{Ferroelectric switching at edge dislocations in BaTiO$_3$\\modelled at the atomic scale}

\author{Himal~Wijekoon}
\affiliation{Interdisciplinary Centre for Advanced Materials Simulation (ICAMS), Ruhr-University Bochum, Universit\"atsstr 150, 44801 Bochum, Germany}%
\affiliation{Center for Interface-Dominated High Performance Materials (ZGH), Ruhr-University Bochum, Universit\"atsstr 150, 44801 Bochum, Germany}
\affiliation{Faculty of Physics and Astronomy, Ruhr-University Bochum, Universit\"atsstr 150, 44801 Bochum, Germany}
\email{himal.wijekoon@rub.de}
\author{Pierre~Hirel}%
\affiliation{Univ. Lille, CNRS, INRAE, Centrale Lille, UMR 8207 - UMET - Unit\'e Mat\'eriaux et Transformations, F-59000 Lille, France}

\author{Anna~Gr\"unebohm}
\affiliation{Interdisciplinary Centre for Advanced Materials Simulation (ICAMS), Ruhr-University Bochum, Universit\"atsstr 150, 44801 Bochum, Germany}%
\affiliation{Center for Interface-Dominated High Performance Materials (ZGH), Ruhr-University Bochum, Universit\"atsstr 150, 44801 Bochum, Germany}%
\affiliation{Faculty of Physics and Astronomy, Ruhr-University Bochum, Universit\"atsstr 150, 44801 Bochum, Germany}
\email{anna.gr\"unebohm@rub.de}

\date{\today}

\begin{abstract}
Ferroelectric switching governs the functional properties of ferroelectric perovskites. It is widely accepted that this switching depends on domain nucleation and pinning and that these processes can be controlled by the defect structure. However, an atomistic picture of the influence of one important class of defects -- dislocations-- on ferroelctric switching is missing. This is an important gap in knowledge as dislocations cannot be avoided at interfaces and can also be engineered by plastic deformation at high temperatures. Using atomistic simulations, we show how the cores of $\langle100\rangle$ edge dislocations in BaTiO$_3$ can either act as nucleation centers for ferroelectric switching or pin walls depending on the direction of the applied field. The coupling between electric field and polarization is strongest when the field is applied parallel to the Burgers vector of the dislocation. 
\end{abstract}

\maketitle


\section{\label{sec:introduction}Introduction}

Ferroelectric perovskite oxides have become critical to a wide range of applications including non-volatile memory devices, actuators and sensors \cite{khosla_integration_2021,veerapandiyan_strategies_2020,whatmore_100_2021,grunebohm_interplay_2021}. Their functional properties  strongly depend on ferroelectric switching and domain wall dynamics \cite{jesse_direct_2008,mokry_identification_2017,fancher_contribution_2017}.
It is established that these dynamics are sensitive to point defects like vacancies or dopants \cite{grunebohm_influence_2023,bulanadi_interplay_2024,feng_defects_2020,gruenebohm_influence_2016}. Depending on type, concentration, and distribution, point defects may act as pinning centers for domain walls, thereby increasing the coercive field (\ec{}), or may act as nucleation centers for switching \cite{sheng_control_2025,mcgilly_dynamics_2017,genenko_mechanisms_2015}. However, domain walls can flow around these 0~D defects and can be unpinned by moderate field strengths, as the active surface area can be small \cite{sheng_control_2025,paruch_nanoscale_2013}. Furthermore, the defect configuration may change with time by defect realignment, diffusion or migration giving rise to aging and functional fatigue \cite{genenko_mechanisms_2015}. 

A larger and time-stable modification of ferroelectric switching and domain walls  has to be expected by the presence of edge-dislocations, which are the lines ($\bm{l}$) at which extra or missing half-planes (along $\bm{b}$)  of atoms end, as illustrated in Fig.~\ref{dislocation_setup}.
These one-dimensional defects induce large strain fields and potentially have a  larger active surface area than point defects. 
After dislocations have been introduced into \bto{} by deformation at high temperatures \cite{doukhan_dislocations_1986,eibl_dislocations_1988,lin_dislocation_2002} or have formed to accommodate interface strain  \cite{nylund_epitaxial_2021,dai_linkup_1996}, they are furthermore  stable and immobile for the typical operation temperatures, even well above the ferroelectric transition temperature. 

In contrast to early predictions based on mesoscopic Landau-Ginzburg-Devonshire theory that ferroelectricity would be suppressed at dislocation cores due to the strain-induced polarization gradients \cite{alpay_can_2004}, it was recently discovered experimentally that engineered dislocation structures can also significantly improve the functional properties of ferroelectrics, i.e.\ enhance dielectric \cite{zhuo_anisotropic_2022} and piezoelectric responses \cite{hofling_control_2021}.  However, experiments do not offer the required spatial and temporal resolution to reveal the microscopic coupling mechanisms. Instead, phase-field simulations have been used to link the observed functional properties to the stabilization and pinning of elastic domain walls at the dislocation cores \cite{zhuo_anisotropic_2022,zhuo_intrinsic-strain_2023}. Furthermore, it has been predicted by phase field simulations that a local enhancement of polarization at dislocation cores with $\vec{b}=[110]$ and $\vec{b}=[100]$ can act as pinning center for ferroelectric $180^{\circ}$ walls \cite{zhou_influence_2022,kontsos_computational_2009,liu_effects_2019}.

While atomistic simulations allowed to explain some of the  properties of dislocations in paraelectric perovskites \cite{hirel_atomistic_2012,marrocchelli_dislocations_2015,klomp_nature_2023,hirel_dislocations_2025}. To the best of our knowledge, only one publication on atomistic modelling of dislocations in ferroelectric \bto{} exists which Deguchi et al.\ reported that dislocation cores can act as nucleation centers for ferroelectric switching and that this process depends on the atomistic core structure and the relative directions of applied field and dislocations using  a machine-learning potential \cite{deguchi_asymmetric_2023}. However, the impact of the cores on the successive switching process has not been addressed so far.  

We present a first atomistic study of the impact of $\langle100\rangle$ edge dislocations in tetragonal \bto{} on the whole  ferroelectric switching process. We show that the field-response around dislocations is highly anisotropic and non-homogeneous.  On the one hand, the dislocations can act as nucleation centers for all field directions with a minimal coupling between the dislocations and polarization along the dislocation line. On the other hand, 180$^{\circ}$ walls are only pinned if the polarization is collinear to the Burgers vector.

\section{\label{sec:methods}Methods}

\begin{figure}[]
    \centering
    \includegraphics[width=.45\textwidth,clip,trim=0cm 0cm 0cm 0cm]{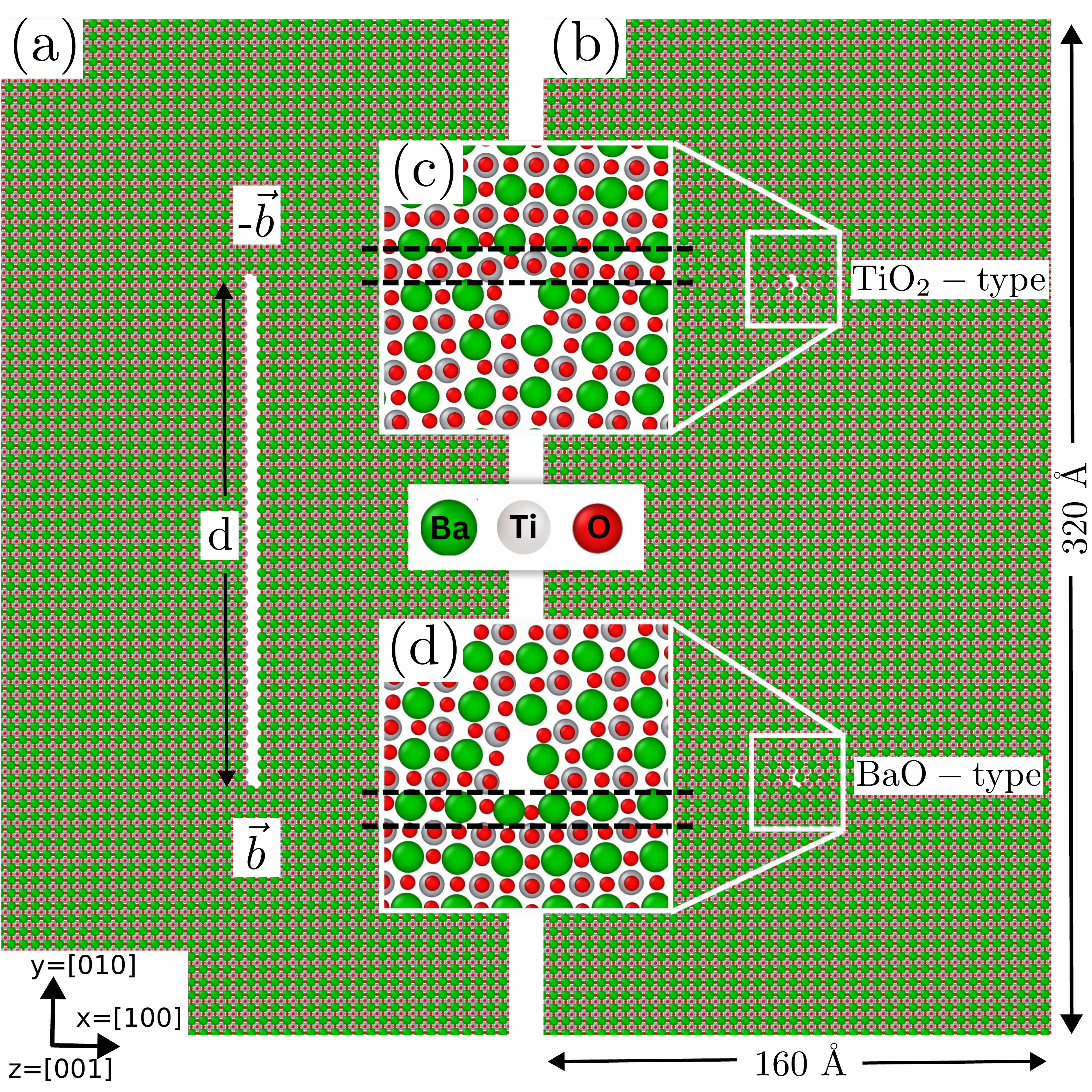}
    \caption{\label{dislocation_setup} Simulation setup: (a) Starting from a $40\times 80 \times 1$ supercell of cubic BaTiO$_3$, a BaO-TiO$_2$ double layer along $[010]$, i.e.\ along y, with a length of $d= 87$~\AA{} has been removed. (b) After relaxation,  a pair of TiO\textsubscript{2}-type and BaO-type dislocation lines along [001], i.e.\ along $z$, with Burger's vectors $b=\pm [100]$ form. (c)-(d) give these cores with atomic resolution.}
\end{figure}

We used the  isotropic, an-harmonic core-shell model parameterized for BaTiO$_3$ with density functional theory simulations by Sepliarsky et al.\ \cite{sepliarsky_atomic_2005}. This potential can correctly reproduce phase stability and domain wall characteristics and dynamics \cite{tinte_ferroelectric_2004,dimou_pinning_2022}. Furthermore, it can correctly reproduce the generalized stacking fault energy and the $(010)$~$\gamma$-surface predicted for tetragonal BaTiO$_3$ by DFT simulations (see Appendix~\ref{coreshell}).

Molecular static simulations were used to relax the structure  with a threshold on the maximal forces on atoms of \SI{e-6}{\electronvolt\per\angstrom} utilizing the LAMMPS package \cite{thompson_lammps_2022} together with the particle-particle-particle mesh  algorithm to treat the Coulomb interaction with a relative accuracy of $10^{-5}$ and a cutoff of 15.0~\AA{}. As the room-temperature tetragonal phase of BaTiO$_3$ is not the energetic ground state at 0~K, we initialized this metastable state in a $10\times 10 \times 10$ unit cells by an  external electric field ($E_z=$\SI{10}{\mega\volt\per\centi\meter}). Starting from the resulting tetragonal configuration, the field hysteresis of the pristine material was recorded with steps of \SI{1}{\mega\volt\per\centi\meter}. At each field increment, atomic positions and volume were relaxed for a constrained cell shape.

To introduce a pair of dislocations under periodic boundary conditions, we used  Atomsk \cite{hirel_atomsk_2015} and adopted the approach reported by Marrocchelli et al~\cite{marrocchelli_dislocations_2015}. A pair of edge dislocations with $\vec{l}$ along $z$-direction, i.e.\ along $[001]$, and $\vec{b}=\pm [100]$, i.e.\ along $\pm x$, was introduced by removing one BaO-TiO$_2$ double layer along $y$-direction from a $40\times 80 \times 1$  cell containing 6320 unit cells of cubic BaTiO$_3$, see Fig.~\ref{dislocation_setup}(a). After that, the cell was compressed by 1~\% along $x$. The tetragonal states along the Cartesian directions $i$, with $i$: $x$, $y$ and $z$, were induced by field poling with $E_i=\SI{100}{\mega\volt\per\centi\meter}$ using the following relaxation protocol. First, the atomic positions were relaxed along $x$-direction, second along $y$-direction, third, a full relaxation of atomic positions, cell shape and volume was performed. Figure~\ref{dislocation_setup}(b) shows the resulting pair of charge neutral edge dislocations and the insets (c) and (d) give the TiO$_2$-terminated dislocation with $\vec{b}=[\Bar{1}00]$ and the BaO-terminated dislocation with $\vec{b}=[100]$ with atomic resolution, respectively. Starting from these relaxed configurations, the field hysteresis was recorded for all three directions $i$. Thereby, a  pair of edge dislocations embedded in a large matrix of the pristine material in its tetragonal phase, was mimicked by constraining the lattice parameters at the cell boundaries to the pristine reference state at the same field strength.  

The local dipole moments, $\bm{u}_k$, of each Ti-centered unit cell, $k$, and the  macroscopic  polarization, $\bm{P}$, were estimated by \cite{sepliarsky_first-principles_2011},
\begin{eqnarray}
    \bm{u}_k &=& \sum_j q_j (\bm{r}_j-\bm{r}_k^{\text{cm}})\,,\\
        \bm{P} &=& \frac{1}{V} \sum {\bm{u}}_k\,,
    \label{dipole_eq}
\end{eqnarray}
with $V$ being the volume of the simulation cell, 
$\bm{r}_k^{\text{cm}}$  being the center of mass of the O-polyhedron  in cell $k$ and with 
$q_j$, $\bm{r}_j$,  being total charges and core positions of all atoms $j$ in the cell $k$, respectively.

To analyze the strain field induced around the dislocation cores, the changes of the Ba--Ba distances along $i$-direction, relative to the pristine tetragonal structure with the same direction of the tetragonal axis, are determined as,
\begin{equation}
 \epsilon_{ii}(k) = \frac{d_{ii}(k) - d_{ii}^0}{d_{ii}^0} 
\end{equation}
where, $d_{ii}(k)$ and $d_{ii}^0$ are the Ba--Ba distances in cell $k$ and without dislocations, respectively.
Ovito was used for all visualizations \cite{stukowski_ovito_2010}.

\section{Results and Discussion}

\begin{figure}[t]
    \centering
    \includegraphics[width=.5\textwidth,clip,trim=0cm 0cm 0cm 0cm]{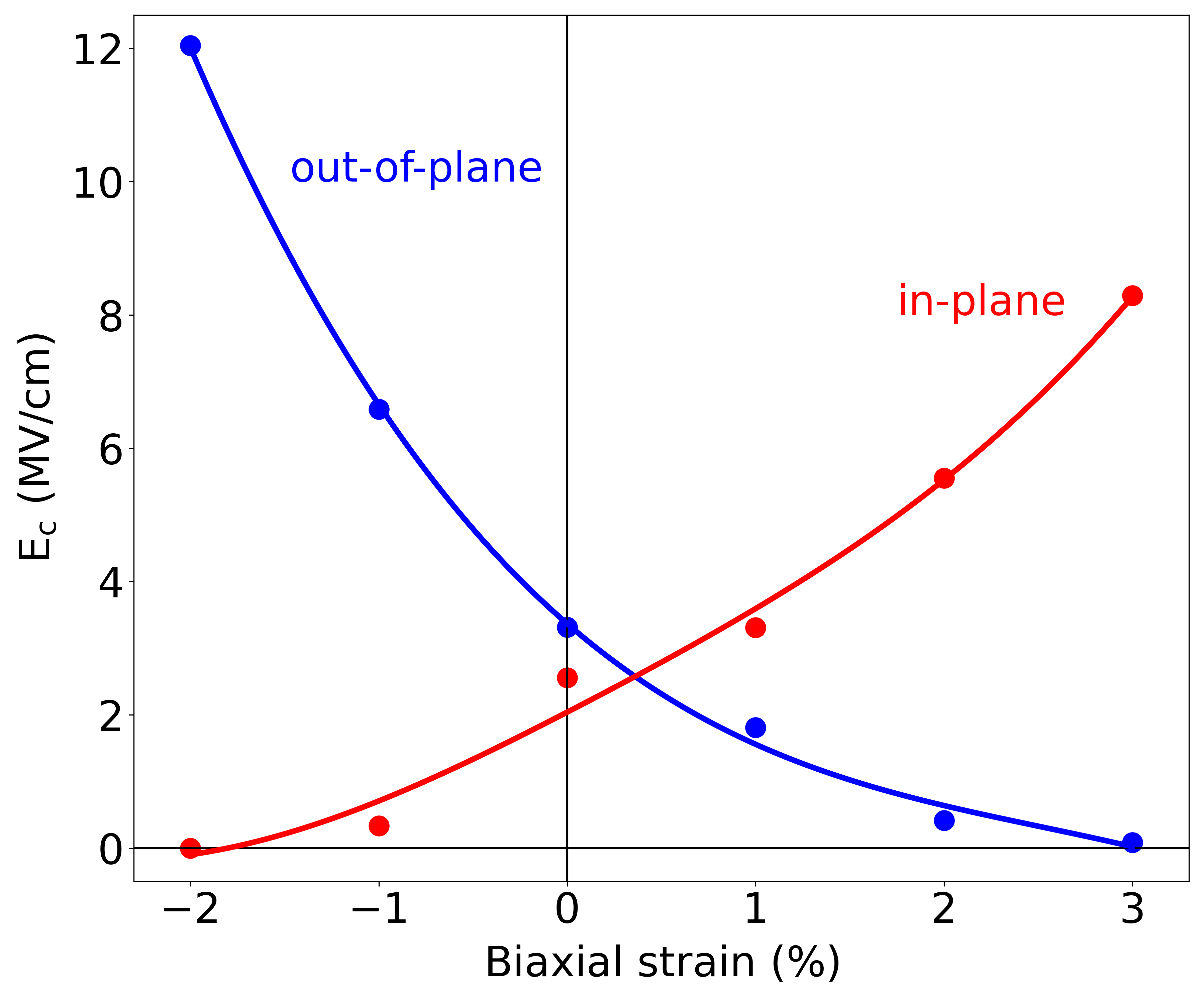}
    \caption{\label{strain_hys_ec} Change of the coercive field ($E_c$) of pristine \bto \ for bi-axial strain in the x-y plane computed using the core-shell model. $\epsilon=0\%$ and $\epsilon=0.2\%$ correspond to the in-plane lattice parameters of the tetragonal phase and of the cubic phase, respectively. Blue and red curves show the coercive fields for the  out-of-plane and in-plane directions. 
  Lines are guides for the eye only.}
\end{figure}

How do dislocations influence ferroelectric switching and field hysteresis? Since one expects the strain field around the dislocation cores to strongly couple to the polarization, it is necessary to first analyze the impact of homogeneous strain on the field-response of pristine tetragonal \bto{}. Without strain, we find $\ec{}=\SI{9}{\mega\volt\per\centi\meter}$ with our simulation protocol. Note that, this strong overestimation of experimental values is expected because of the absence of nucleation centers and thermal excitation \cite{landauer_electrostatic_1957}. Furthermore, the imposed constraints on the tetragonal ratio enhances the \ec{}. If the tetragonal axis is relaxed for a clamped film with $\epsilon=0$~\%, the \ec{} is reduced to \SI{3}{\mega\volt\per\centi\meter}. The blue curve in Figure~\ref{strain_hys_ec} illustrates the change of \ec{} under biaxial-strain in the $xy$-plane and the field applied along $z$-direction. On the one hand,  the polarization along $z$-direction and the \ec{} along this out-of plane direction are reduced with tensile strain. For $\epsilon=3$\%, the out-of plane polarization is no longer stable without the applied field. On the other hand, the out-of plane polarization and \ec{} increases about 50\% for 1\% compressive strain increment, in agreement to literature \cite{choudhury_strain_2008,azuma_tuning_2023}.

The opposite trend is found if the field is applied along $x$ (or $y$)-direction, i.e.\ in the strained plane (red curve in Fig.~\ref{strain_hys_ec}). In these cases, \ec{} increases and decreases under tensile and compressive strain, respectively. However, as both in-plane lattice parameters are equal, the symmetry of the system is not tetragonal for all values of strain. The change of \ec{} under tensile strain is smaller than change of \ec{} under compression and in-plane polarization is no longer stable for strain already exceeding $\epsilon=-2$\%. In summary, the direction and magnitude of ferroelectric polarization and particularly the \ec{} are very sensitive to magnitude and direction of strain. \\

\begin{figure}[t]
    \centering
    \includegraphics[width=.5\textwidth,clip,trim=0cm 0cm 0cm 5cm]{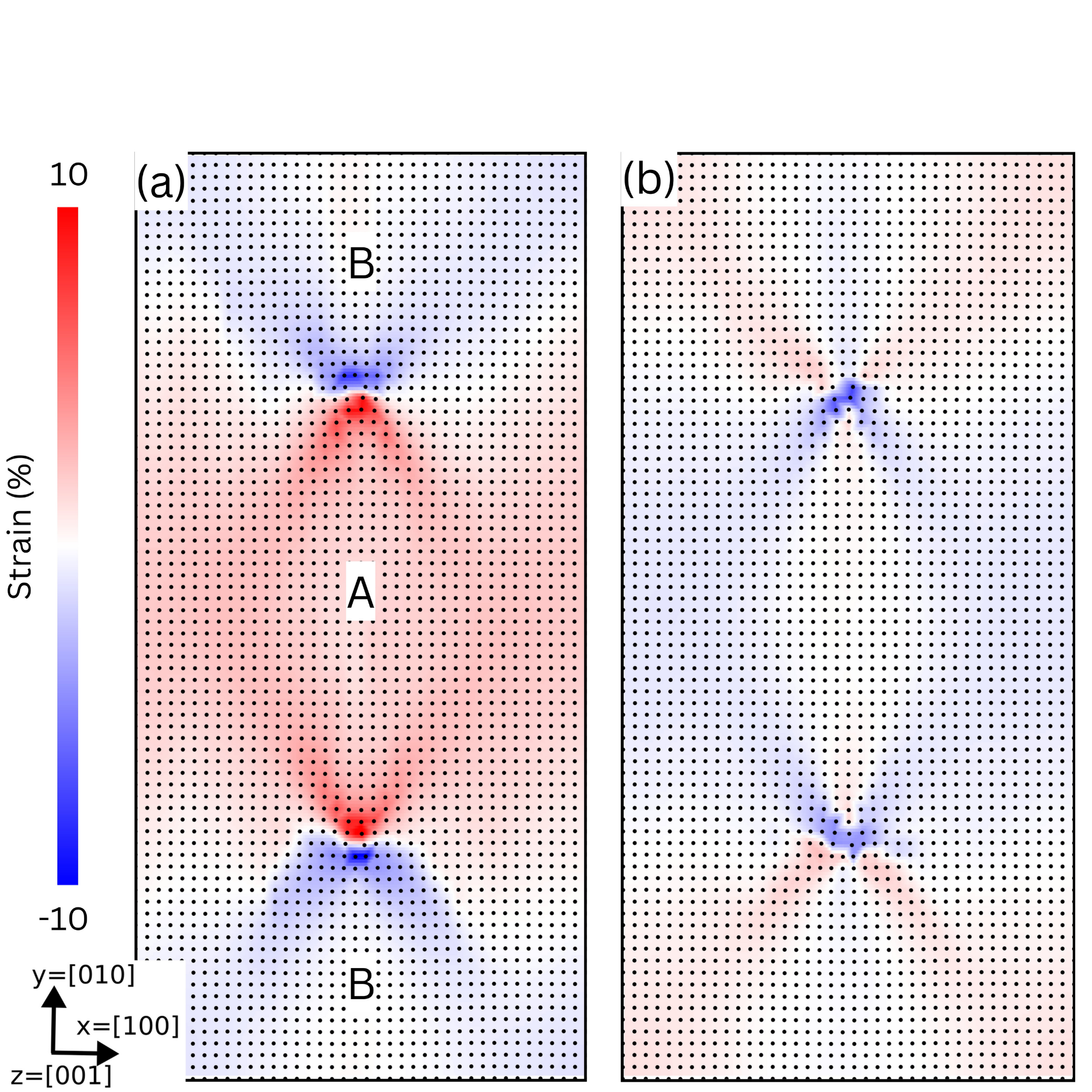}
    \caption{\label{induced_dipoles_strain} (a)--(b) Strain map around the pair of edge dislocations shown in Fig.~\ref{dislocation_setup} in the tetragonal phase polarized along $x$.
    The strain along (a) $x$ ($\epsilon_{xx}$) and  (b)  $y$ ($\epsilon_{yy}$) directions are given relative to the pristine material with tetragonal axis along $x$-direction.  Ba atoms are given as spheres and red or blue colour indicate  tensile or compressive strains, respectively.}
\end{figure}

Figure~\ref{induced_dipoles_strain} shows the strain field induced by the pair of edge dislocations in the tetragonal phase polarized along $x$-direction, i.e.\ collinear to $\vec{b}$. Note that, despite the differences in composition (Ba-rich vs.\ Ti-rich) and resulting atomic relaxation at both cores, see insets in Fig.~\ref{dislocation_setup}, the strain field is approximately symmetric. In agreement to macroscopic elastic theory \cite{hirth_book_1982}, the dislocations with $\vec{b}=\pm[100]$ induce $\epsilon_{xx}>0$ and $\epsilon_{xx}<0$ in region A and region B, respectively, and vice versa for $\epsilon_{yy}$. Thereby, the magnitude of strain is maximal along $\langle 110 ]$ and exceeds $\epsilon_{xx}=10.0$~\%  and $\epsilon_{yy}=-3.5$~\% in about four unit cells next to the cores. Locally, the strain along $x$ and $y$ directions exceed the values of bi-axial strain discussed above. One may thus expect large gradients of the \ec{} along $x$ and $y$ directions closer to the cores. 

The strain maps for the tetragonal axes aligned along $y$ and $z$ directions, i.e.\ perpendicular to $\vec{b}$, are given in Appendix~\ref{strain_profiles}. For these orientations, $\epsilon_{xx}>0$ is induced in the whole volume A with the missing double layer, while the volume with $\epsilon_{xx}<0$ is restricted to the $\langle 110 \rangle$ wedges next to the cores in region B. Furthermore, the lattice is compressed along the polarization direction $P_y$ ($\epsilon_{yy}\geq -3$~\% at the core). Compared to $P_x$, the area with sizeable compression at the wedges in region A is considerably larger and slightly anisotropic, (Appendix~\ref{strain_profiles}~(d)). For the chosen boundary conditions, $\epsilon_{zz}$ is zero per definition for all polarization directions. Also local variations are not induced by infinitely long straight dislocation lines along that direction.

\begin{figure}[h]
    \centering
    \includegraphics[width=0.48\textwidth,clip, trim=0cm 0cm 0cm 0cm]{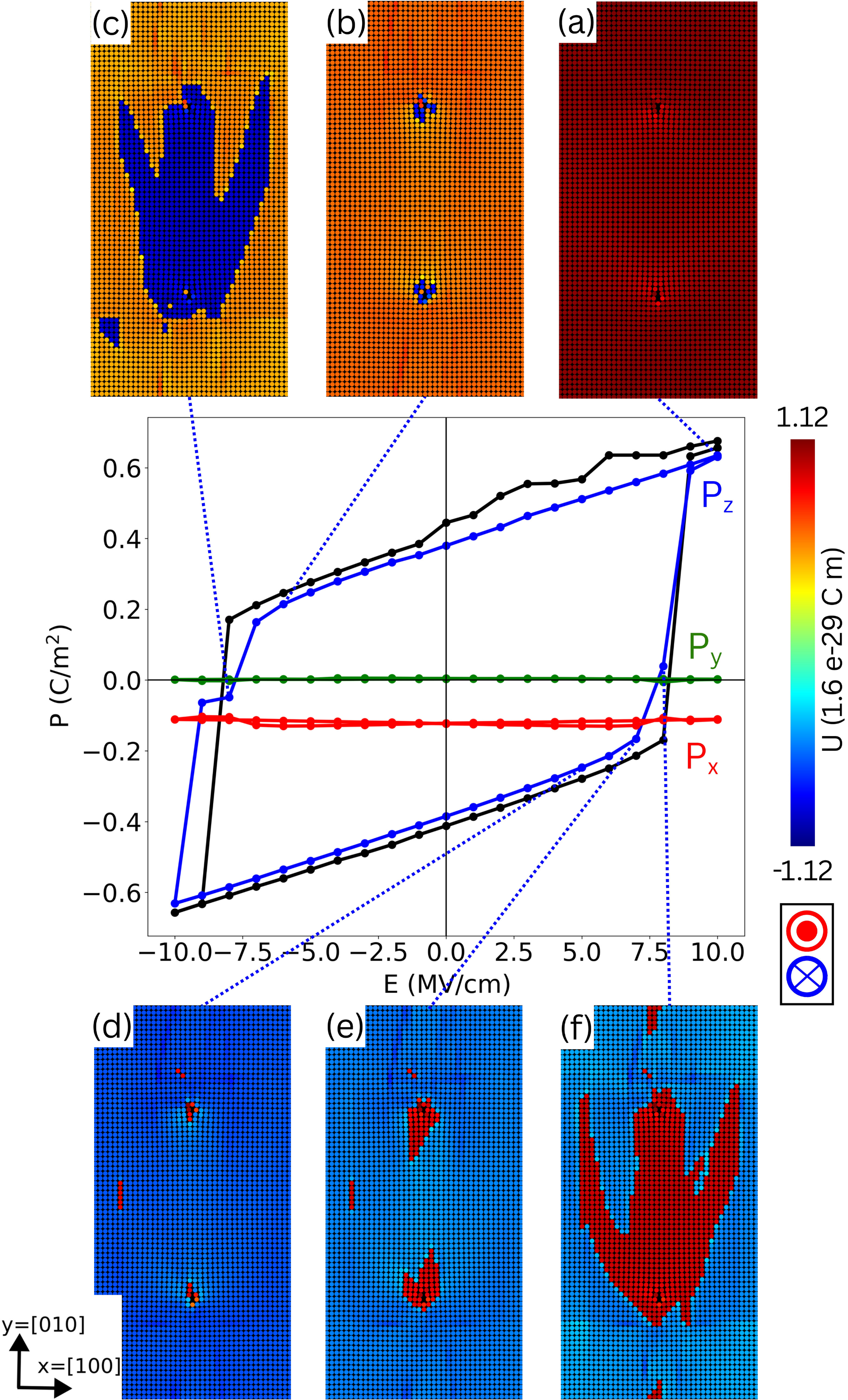}
    \caption{\label{pz_switching} Field hysteresis for an electric field applied along $z$-direction in the presence of a pair of dislocations with $\vec{b}$ and $\vec{l}$ along $x$ and $z$ directions, respectively. Colors mark the three polarization components and the hysteresis of the pristine material is given in black. 
    (a)--(f) Underlying microscopic processes color encoded by the  magnitude of the local polarization with red and blue indicating $P_z$ and $-P_z$, respectively.}
\end{figure}

To test how these strain fields influence the field-induced switching, we start with the case of the polarization in an electric field applied parallel to the dislocation lines ($E_z$). Figure~\ref{pz_switching} shows the evolution of the polarization with the field strength. The macroscopic polarization along the field direction $P_z$ (blue) and the related \ec{} are reduced by about 5\% and 8\%, respectively. This small impact of the dislocations on the field hysteresis can be understood by the absence of strain along $z$-direction. Furthermore, a macroscopic monoclinic polarization rotation with $P_x=\SI{-0.12}{\coulomb\per\meter}$ (red) is induced by local polarization rotation towards $x$-direction in region A with tensile $\epsilon_{xx}$. 

Figures~\ref{pz_switching}(a)--(f) show snapshots of the local dipole moments ($u_z$). In the positively poled state (Fig.~\ref{pz_switching}(a)), $P_z$ is mainly homogeneous with a small reduction around the cores. If the field is reversed, the switching starts independently at both cores, see Fig.~\ref{pz_switching}(b). Between \SI{6}{\mega\volt\per\centi\meter} and \SI{10}{\mega\volt\per\centi\meter}, first region A and then region B switch. The system is homogeneous along $z$-direction and thus no domain walls perpendicular to the $xy$-plane form. The interfaces between $\pm u_z$ in the plane do not induce polarization gradients along $z$-direction and have local $\langle 110\rangle$, $[100]$ and irregular normales. The microscopic switching processes are reversible (Fig.~\ref{pz_switching}(d)--(f)).  Note that a few dipoles do not switch back, which could be an artefact of the relaxation procedure. Without thermal excitation, possible changes of \ec{} compared to the pristine material, are below the field resolution. However at finite temperatures, the nucleation at the cores may be more relevant. \\

\begin{figure}[t]
    \centering
    \includegraphics[width=0.48\textwidth,clip, trim=0cm 0cm 0cm 0cm]{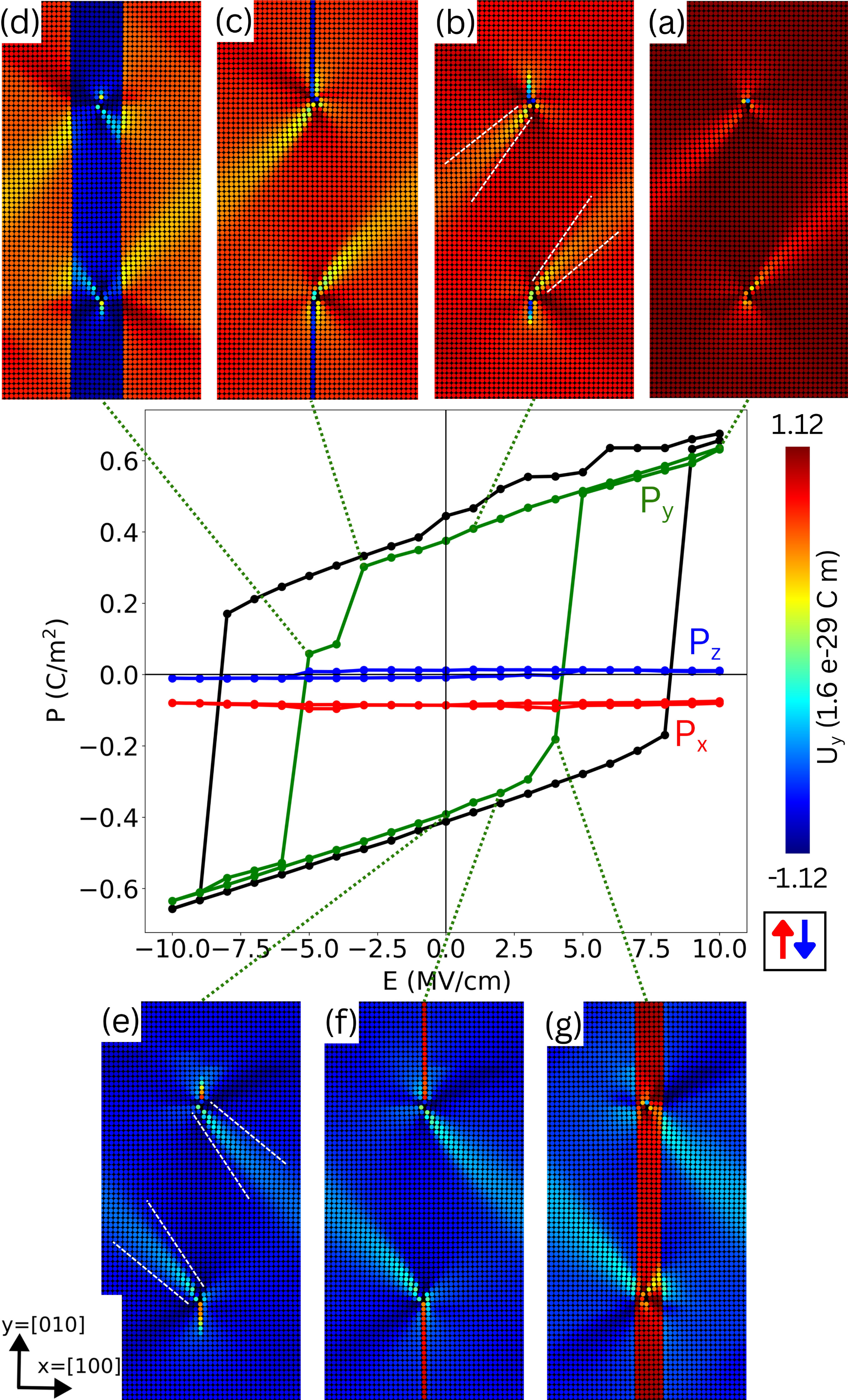}
    \caption{\label{py_switching} Field hystereses for an electric field applied along $y$ in the presence of a pair of dislocation with $\vec{b}$ and $\vec{l}$ along x and z. Colors mark the three polarization components  and the hysteresis of the pristine material is given in black. 
    (a)--(f) Underlying microscopic processes color encoded by the  magnitude of the local polarization with red and blue indicating $P_y$ and $-P_y$.}
\end{figure}

As discussed above, larger strain variations are induced perpendicular to the dislocation line $\vec{l}$. Therefore the dislocations induce larger changes of the $E_y$ hysteresis, i.e.\ for the field applied perpendicular to $\vec{l}$ and $\vec{b}$ (Fig.~\ref{py_switching}). While the magnitude of the macroscopic saturation polarization is close to that of the pristine material, \ec{} is reduced by about 40~\%. Again, the tensile strain $\epsilon_{xx}$ in region A (see Fig.~\ref{induced_strain_supp}(b)) induces local polarization of up to $P_x=\SI{0.223}{\coulomb\per\meter\squared}$ which results in a mean polarization of $P_x=\SI{-0.08}{\coulomb\per\meter\squared}$ for the chosen system size. 

Figures~\ref{py_switching}~(a)--(g) show the evolution of the local dipoles $u_y$ with the applied field. Even under an applied field of $E_y=$\SI{1}{\mega\volt\per\centi\meter} the polarization $P_y$ is locally reduced due to the local strain $\epsilon_{yy}<0$ particularly along $\pm[110]$ in region A. In line with the strain field, the local reduction is maximal on $\pm [110]$-wedges in region A (white dashed-lines in Fig.~\ref{py_switching}(b)). For $E_y=$ \SI{2}{\mega\volt\per\centi\meter}, the single-domain ferroelectric state polarized along $y$-direction is no longer stabilized by the field. Instead, about 4 dipoles at the cores are reversed as discussed below. With increasing negative field, the polarization switching starts at these clusters with the formation of a needle-like domain with a width of one unit cell which spans the whole region B (Fig.~\ref{py_switching}(c)). This needle-shape is related to the unfavourable depolarization field at head-to-head and tail-to-tail interfaces of a switched cluster along $y$-direction \cite{khachaturyan_domain_2022}. Second, this needle expands along $x$-direction and penetrates region B along $y$-direction (\SI{-3}{\mega\volt\per\centi\meter} -- \SI{-5}{\mega\volt\per\centi\meter}) resulting in an intermediate state with two 180$^{\circ}$ domain walls parallel to the field direction. The critical field to shift these domain walls through the system and complete the switching is about $E_y=$ \SI{8}{\mega\volt\per\centi\meter} and \ec{} is thus reduced by about 33\% compared to the pristine material. This can be understood as the critical field to shift existing domain walls is considerably smaller compared to that for homogeneous switching, which is also holds true for the pristine material \cite{khachaturyan_domain_2022,boddu_molecular_2017}. Note that, there is also no large energy barrier to shift these walls as $\epsilon_{yy}$ is reduced compared to pristine material (Fig.~\ref{induced_strain_supp}(b) in Appendix~\ref{strain_profiles}). Again, the switching is reversible. In the negative field, however $\epsilon_{yy}<0$ and $P_z$ are reduced in region A on the wedges which are along $\pm[\bar{1}10]$. \\

\begin{figure}[t]
    \centering
   \includegraphics[width=0.48\textwidth,clip, trim=0cm 0cm 0cm 0cm]{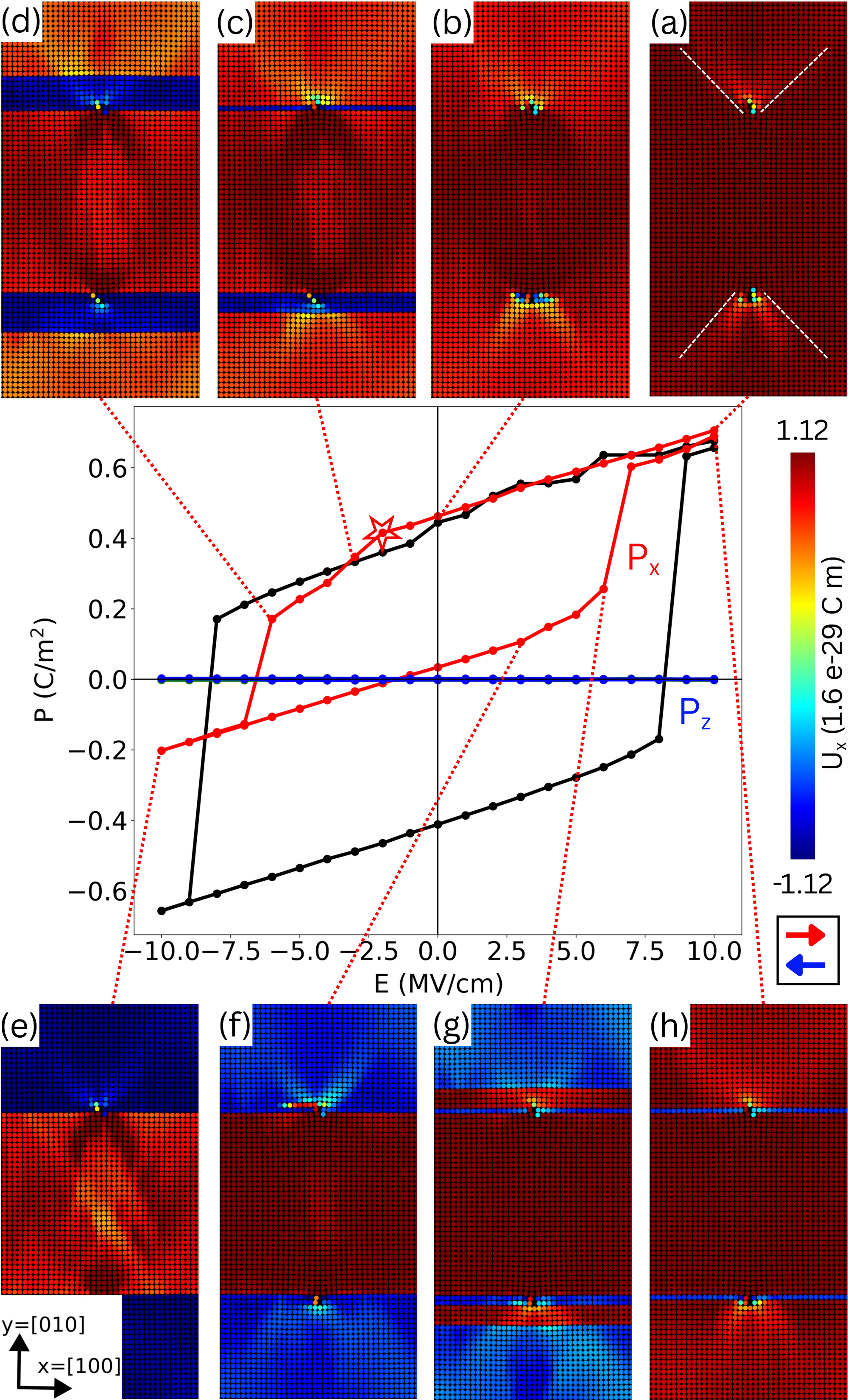}
    \caption{\label{px_switching}
Field hystereses for an electric field applied along $x$ in the presence of a pair of dislocation with $\vec{b}$ and $\vec{l}$ along x and z. Colors mark the three polarization components  and the hysteresis of the pristine material is given in black. 
    (a)--(f) Underlying microscopic processes color encoded by the  magnitude of the local polarization with red and blue indicating $P_x$ and $-P_x$.   } 
\end{figure}

The impact of the dislocations on strain, polarization and ferroelectric switching is maximal if the tetragonal axis is parallel to the Burgers vectors, i.e.\ for $E_x$, see Fig.~\ref{px_switching}. In this case, the hysteresis loop is highly asymmetrical and no macroscopic monoclinic distortion is induced. After field poling (Fig.~\ref{px_switching}(a)--(b)), the system is in a single-domain ferroelectric state with \pr{} equals to that of the pristine material. Without an external field, the polarization in region B with $\epsilon_{xx}<0$ is however reduced compared to the pristine material (by up to about 93\% directly at the cores) while the tensile $\epsilon_{xx}$ in region A induces an increase of the local polarization $u_x$. 

Analogous to $P_y$, the switching of $P_x$ sets in by the formation of needle-like domains spanning the whole system along the polarization direction at $E_x=\SI{-1}{\mega\volt\per\centi\meter}$. With increasing field strength, these domains grow by the shift of the walls along $\pm y$-direction through region B (Fig.~\ref{px_switching}(c)--(d)). At about $E_x=\SI{-7}{\mega\volt\per\centi\meter}$, i.e.\ below \ec{} of the pristine material, the whole region B has switched and the polarization is zero for equal dislocation spacing in region A and B. However, even the field of $E_x=\SI{-10}{\mega\volt\centi\meter}$ is not sufficient to shift the walls through the layer with $\epsilon_{xx}$ up to about 16\% at the border of region A. Both walls are pinned directly behind the core (after the first full Ti-plane in  Fig.~\ref{px_switching}(e)).
Furthermore, there is no onset of switching in region A, analogous to the discussed increase of \ec{} of the strained pristine material, as $\epsilon_{xx}>1$~\%  in the whole region. With increasing field strength, only the polarization magnitude in the domains parallel or anti-parallel to the field increase or decrease, respectively, resulting in a reversible, linear change of $P_x$ with $E_x$. 

If the field is reversed, new domains parallel to the field direction nucleate at $E_x=\SI{3}{\mega\volt\per\centi\meter}$ in region B where the compression along $x$-direction is maximum (Fig.~\ref{px_switching}(f)). These domains are two layers apart from the persisting positive domain. 
With  increasing field strength, the domain walls shift through region B with $\epsilon_{xx}<0$ and at $E_x=\SI{10}{\mega\volt\centi\meter}$ switching completes (Fig.~\ref{px_switching}(h)). The pinned planes at the core (Fig.~\ref{px_switching}(h)) result in a reduction of the saturation polarization in successive cycles. Qualitatively, the same trends--the nucleation of the polarization switching at the dislocation cores, followed by shifting of domain wall for $E_x$ and $E_y$ or abrupt switching in case of $E_z$ are observed, if the constraints on the cell shape are lifted and the simulation cell is fully relaxed during the field cycle.

\begin{table}[]
\caption{The coercive field ($\ec$) saturation polarization ($\pr$) for the different relative directions of applied field and dislocations compared to pristine BaTiO$_3$.}
\label{table:hys}
\begin{ruledtabular}
\begin{tabular}{l|lcc}
&Configuration & $\ec$  & $\pr$ \\
&& (\si{\mega\volt\per\centi\meter}) & (\si{\coulomb\per\meter\squared}) \\
\hline
\multicolumn{2}{c}{Pristine }
& 9.0 & 0.41   \\
$E_x$&$\vec{E}\parallel\vec{b}$
& 7.0 & 0.43   \\
$E_y$&$\vec{E}\perp\vec{l},\vec{b}$ 
& 6.0 & 0.395  \\
$E_z$&$\vec{E}\parallel\vec{l}$ 
& 8.0 & 0.39  \\
\end{tabular}
\end{ruledtabular}
\end{table}

To summarize, the characteristic \ec{} and \pr{} for each relative direction between dislocation and field are compared to the pristine material in Tab.~\ref{table:hys}. Note that, the present results are not related to thermally activated nucleation processes as we discuss the limiting case of switching of the fully relaxed system. In the presence of thermal fluctuations, the reduced energy barriers for switching can already be crossed for smaller critical field strengths. As one can understand by the strain fields, the critical field for the nucleation of the switching process is smallest for the field applied perpendicular to Burgers vector and dislocation line, while the impact of dislocations is minimum if the field is applied parallel to the Burgers vector. While the onset of switching and the \ec{} are also reduced by a field parallel to $\vec{b}$, where one half of the system switches in the chosen field interval.\\

\begin{figure}[t]
    \centering
    \includegraphics[width=.5\textwidth,clip, trim=0cm 0cm 0cm 0cm]{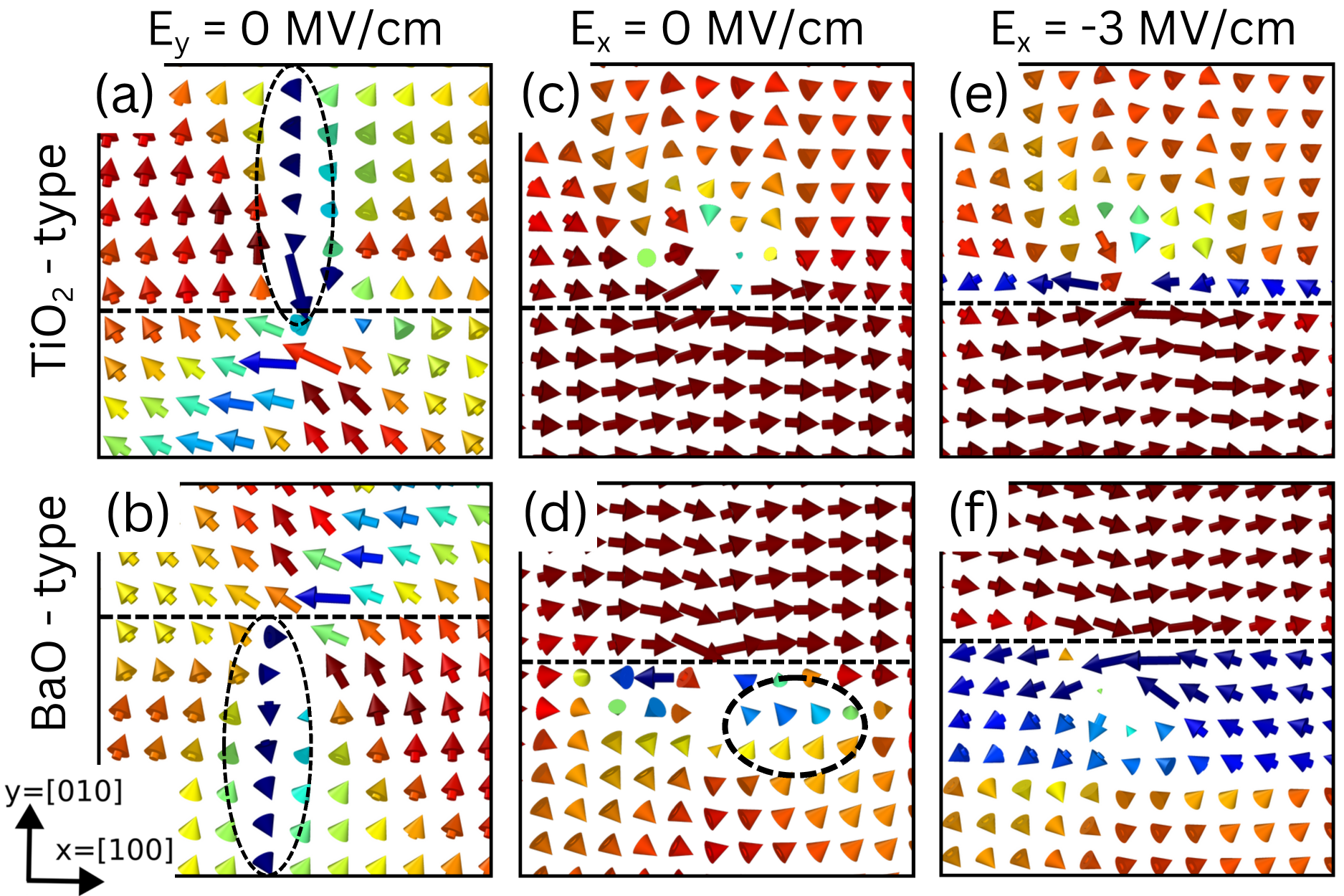}
    \caption{\label{switching_zooomed} Local polarization at the (top: TiO$_2$-type and bottom: BaO-type) dislocation cores with unit cell resolution after field poling along (a)--(b) $y$ and (c)--(d) $x$ directions and after the negative field was increased to $E_x=\SI{-3}{\mega\volt\centi\meter}$. Straight dashed lines separate regions A and B.} 
\end{figure}

To gain a microscopic understanding of the initial nucleation processes, figures~\ref{switching_zooomed}~(a)--(d) show both cores after field poling with unit cell resolution. After poling along $y$-direction (Fig.~\ref{switching_zooomed}~(a)--(b)), the local dipoles apart from the cores point along $y$-direction. In the strain profiles around both cores, the local polarization however shows the discussed rotation towards $x$-direction ($-u_x$) in region A. As this polarization rotation induces a gradient of $P_y$ along $y$-direction, $P_y$ is locally reduced and one row of dipoles switched the direction to reduce the depolarization field in region B next to both cores.

After field poling along $x$-direction (Fig.~\ref{switching_zooomed}~(c)--(d)), the local dipoles apart from the cores and in region A point along $x$-direction. Although $P_y$ is strongly reduced by the strain field next to the core in region B, this does not induce a polarization gradient along $y$. Therefore, no full line of reversed dipoles exists which could promote polarization switching in the reversed field. The different relative direction of field and $\vec{b}$ can thus partly explain the smaller critical field strength for nucleation for $E_y$ compared to $E_x$. 
Furthermore, Figures.~\ref{switching_zooomed}~(c)--(d) show that the switching from $x$ to $-x$-directions, sets in at the BaO-type core first. When the field is further increased to $E_x=\SI{-3}{\mega\volt\per\centi\meter}$, the switched volumen at the BaO-centered core increases and  the plane next to the TiO$_2$-terminated core switches.

\section{Summary and Conclusions}
In this work we analyzed the impact of $\langle100\rangle$ edge dislocations on field-induced switching in tetragonal BaTiO$_3$ by means of atomistic simulations. In contrast to the common belief that dislocations act as pinning centers for domain walls, we have demonstrated that the dislocation cores can act as energetically favourable domain nucleation sites. Thereby, the reduction of the coercive field is maximal, if the field is applied perpendicular to Burgers vector and dislocation line and may be sensitive to the termination of the core. Furthermore, we have shown that dislocation cores pin non-elastic 180$^{\circ}$ domain walls by their compressive strain field which may reduce the switchable polarization, only if the field is applied along the Burgers vector.

\section*{acknowledgments}
We acknowledge financial support by the DFG Emmy-Noehter group:  412303109

\section*{Data Availability Statement}
The data that support the findings of this study are available from the corresponding author upon reasonable request. 

\bibliography{refs}

\clearpage
\appendix
\renewcommand{\thefigure}{\thesection\arabic{figure}}
\setcounter{figure}{0}

\section{Core-shell potential}
\label{coreshell}
We used the isotropic core-shell model from \cite{sepliarsky_atomic_2005} presented in table \ref{tableS1}. Each atom is represented by a core and shell with fractional charges, which are connected by an anharmonic spring interaction. The core-shell coupling potential $V^{cs}(w)$ is given by,
\begin{equation}
    V^{cs}(w) = \frac{k_2w^2}{2} + \frac{k_4w^4}{24}
\end{equation}
where, $w$ is the distance between core and shell within an ion. Short-range interactions are only considered between shells and described by a Buckingham potential $V^{B}(d)$, while long-range Coulombic interactions $V^{c}(r)$ act between both cores and shells.
\begin{equation}
    V^{B}(d) = Aexp(\frac{-d}{\rho}-\frac{C}{d^6})
\end{equation}

\begin{equation}
    V^{c}(r) = \frac{1}{4\pi\epsilon_0} \sum_{j\not=i}^N \frac{q_iq_j}{r}\;, 
\end{equation}
where, $q_i$ and $q_j$ are the fractional charges of particles $i$ and $j$.

\begin{table}[h]
\caption{\label{tableS1} Core-shell potential parameters used in this study taken from \cite{sepliarsky_atomic_2005}. For all parameters, energy, length and charge are given in eV, \AA{} and electrons, respectively.}
\begin{ruledtabular}
\begin{tabular}{lcccc}
Atom & Core charge & Shell charge & $k_2$ & $k_4$  \\
\hline
Ba & 5.62 & -3.76 & 251.8 & 0.0  \\
Ti & 4.76 & -1.58 & 322.0 & 500.0 \\
O & 0.91 & -2.59 & 31.0 & 4000.0 \\
\hline 
Short range & $A$ & $\rho$ & $C$\\
 Ba-O & 1061.30 & 0.3740 & 0.0\\
 Ti-O & 3769.93 & 0.2589 & 0.0 \\
 O-O & 4740.00 & 0.2686 & 160.0 
\end{tabular}
\end{ruledtabular}
\end{table}

\begin{figure}[h]
    \centering
    \includegraphics[width=.5\textwidth, clip, trim=0cm 0cm 0cm 0cm]{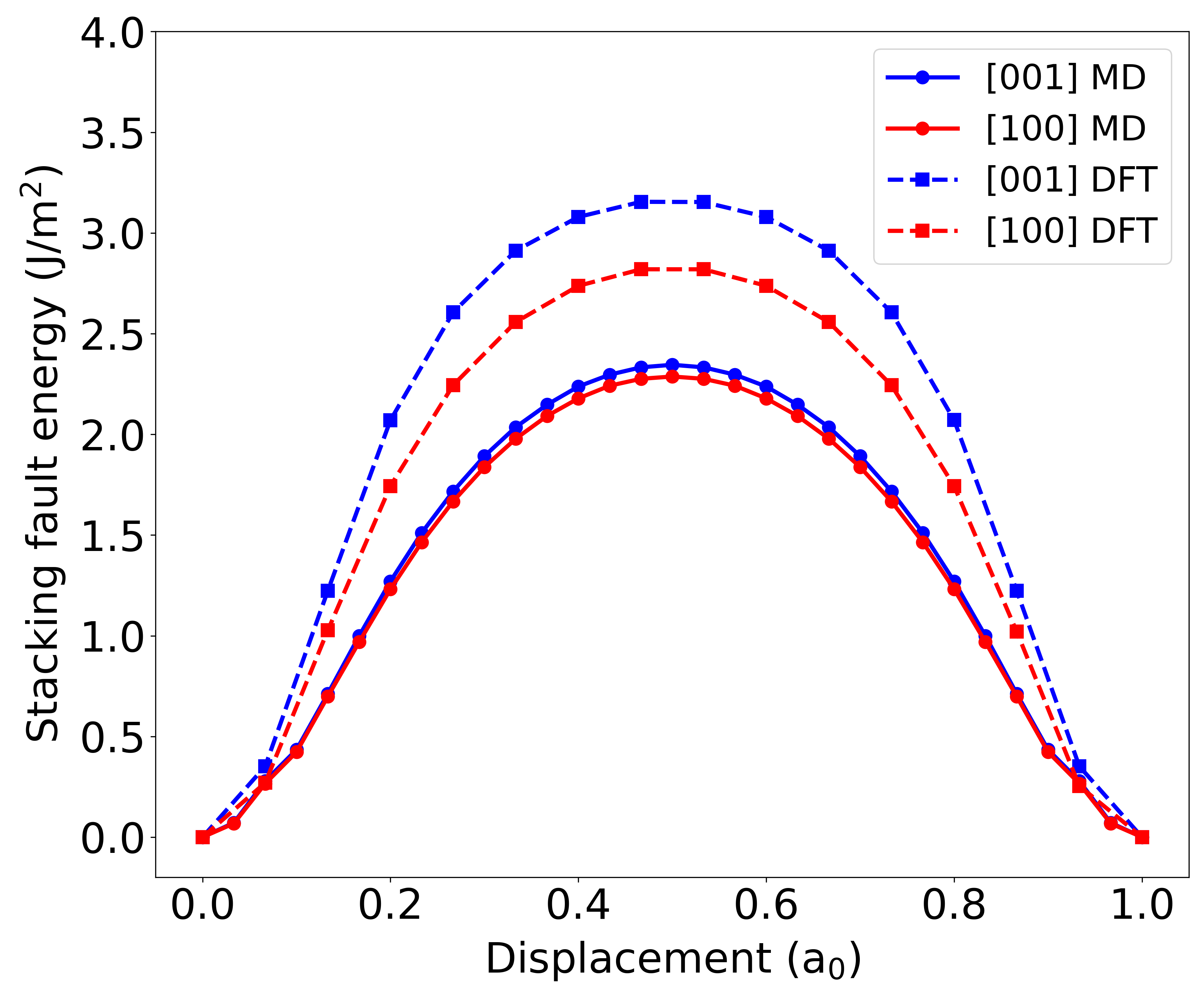}
    \caption{\label{tBTO_010_gamma_2d_surface} (a) [001] and [100] energy profiles of $(010)$~$\gamma$-surface in tetragonal phase with polarization along [001], computed using core-shell potential and PBE.} 
\end{figure}

In order to validate the core-shell potential, Table~\ref{tableS2} compares the predicted lattice parameters to literature. 
 Furthermore, we also computed stacking fault energy profiles along [001] and [100] of teragonal BaTiO$_3$ polarized along [001], see Fig.~\ref{tBTO_010_gamma_2d_surface}. The qualitative trends of the stacking fault and the lattice parameters are in qualitative agreement to DFT predictions. However, the stacking fault energy is quantitatively underestimated. 

\begin{table}[h]
\caption{\label{tableS2} 
Structural properties of tetragonal BaTiO$_3$ calculated in this study compared with DFT values.}
\begin{ruledtabular}
\begin{tabular}{lcc}
Property & This study & Literature \\
\hline
a (\AA{}) & 3.959 & 3.994 \cite{ghosez_first_1999} \\
c (\AA{}) & 4.112 & 4.036 \cite{ghosez_first_1999}  \\
APB (cubic phase), $\gamma$~(J$\cdot$m$^{-2}$) & 1.215 & 0.929 \cite{hirel_theoretical_2010}
\end{tabular}
\end{ruledtabular}
\end{table}

\clearpage
\section{Strain profile of $y$ and $z$-poled systems}
\label{strain_profiles}

\begin{figure*}[h]
    \centering
    \includegraphics[width=\textwidth,clip,trim=0cm 0cm 0cm 0cm]{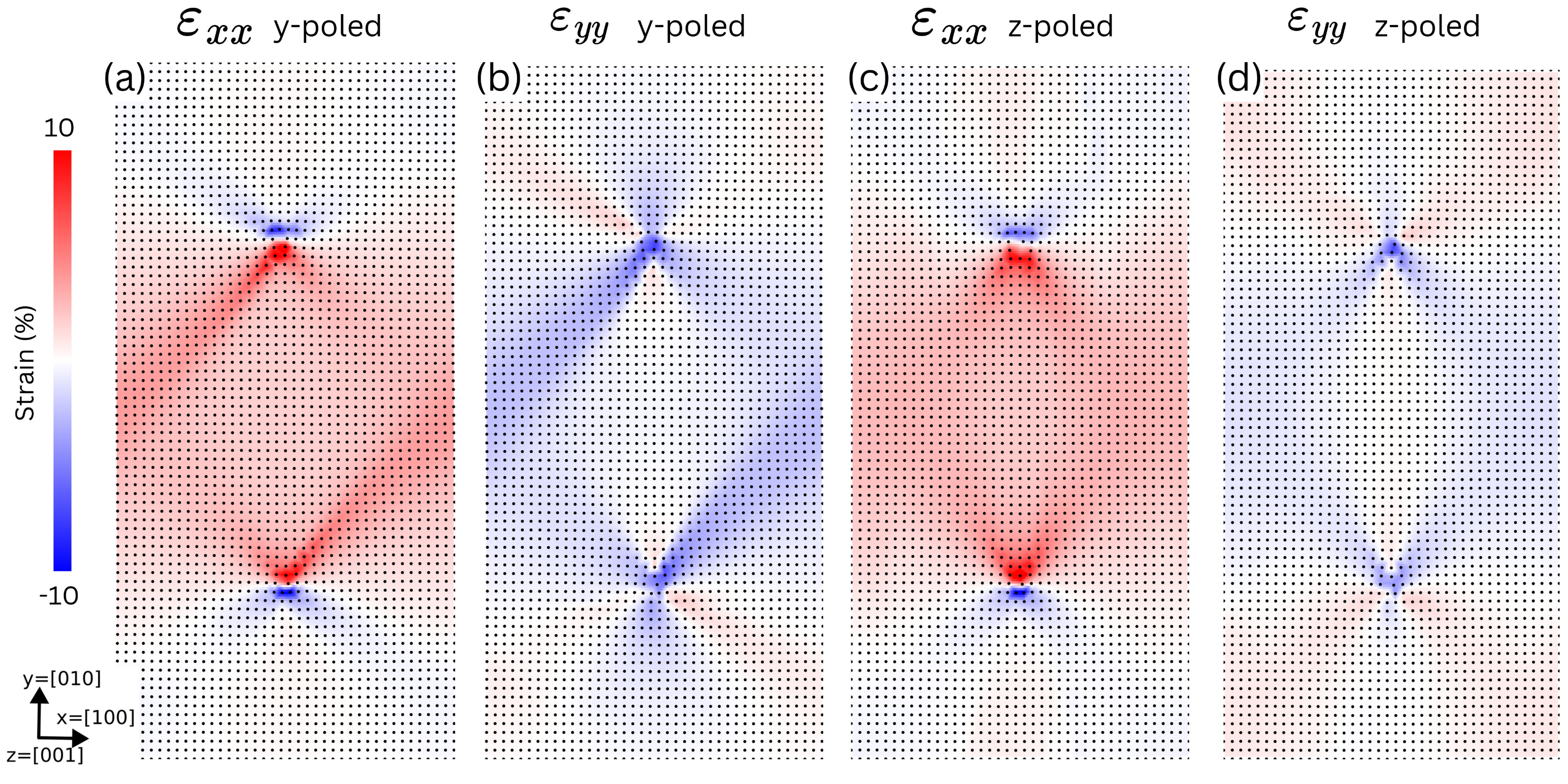}
    \caption{\label{induced_strain_supp} Strain maps around the pair of edge dislocations shown in Fig.~\ref{dislocation_setup} in the tetragonal phase polarized along (a)--(b) $y$-direction and (c)--(d) $z$-direction. The strains are given relative to the pristine material. Ba atoms are given as black dots and red or blue colour indicate tensile or compressive strains, respectively.}
\end{figure*}

Figure~\ref{induced_strain_supp} shows the strain fields induced around the pair of dislocations after poling along fig.~\ref{induced_strain_supp}(a)--(b) $y$-direction and figure~\ref{induced_strain_supp}(c)--(d) $z$-direction. Analogous to the field poling along $x$-direction discussed in the main paper, the dislocations induce $\epsilon_{xx}>0$ and $\epsilon_{xx}<0$ in region $A$  and region $B$, respectively, and vice versa for $\epsilon_{yy}$ for both field directions. The tensile $\epsilon_{xx}$ and compressive $\epsilon_{yy}$ strains are maximum along the wedges of $\pm [110]$ for system poled in $y$-direction, see fig.~\ref{induced_strain_supp}(a)--(b). Strain difference of these wedges and other regions prominent for this case compared to other two. For all applied field directions, dislocations do not induce strain along $z$-direction. Hence, strain maps along $z$-direction are not shown.

\end{document}